# Systematic teaching of UML and behavioral diagrams

**Anja Metzner**
*(Technical University of Applied Sciences Augsburg, Faculty of Computer Science,* Augsburg, Germany, anja.metzner@tha.de*)*

*Abstract*— When studying software engineering, learning to create UML diagrams is crucial. Similar to how an architect would never build a house without a building plan, designing software architectures is important for developing high-quality software. UML diagrams are a standardized notation for the visualization of software architectures and software behavior.

**The research question that inspired this work was how to effectively evaluate hand-drawn diagrams without relying on model parsers. The findings of this investigation are presented in this paper. This article discusses the systematic acquisition of skills required for creating UML diagrams. Especially well-formed activity diagrams are one highlight.**

**Additionally, the paper provides a variety of exercises. The exercises use recommended question types. The more unusual question types are related to images, such as questions about image annotation, finding hotspots on an image and positioning a target on an image. All the demonstrated exercises are suitable for both digital and handwritten training or exams.** (*Abstract*)

*Keywords*— *UML modeling, systematic, competence-based teaching, well-formed activity diagrams (key words)*

## I. INTRODUCTION

Are you teaching modeling with UML [1]? Are you responsible for evaluating UML diagrams created by students? How can we professionals evaluate drafted activity diagrams fairly and grade appropriately? How can we effectively train and assess student comprehension of sequence diagrams? Software engineering teachers are confronted with questions similar to those above. Their goal is to design appropriate trainings and exams.

This paper proposes strategies for improving the expertise of students in creating UML diagrams through the use of a step-by-step approach. I propose a systematic and reliable process for proofreading and assessing students' diagrams without using parsers (e.g. [17, 18, 8, 9]). Suitable examples of exercises are provided for demonstrating the strategies.

### A. Methodology

Teaching UML modeling is challenging. One challenge is the broad applicability of UML beyond software engineering. Thus, this work concentrates on software modeling. In the context of Germany's educational landscape, I outline a competency-based approach for teaching UML. Additionally, I address controversial aspects of standard UML notation and propose solutions. The rest of the paper introduces an approach to teaching UML through relevant exercises.

*1. Related Work*

This paper primarily focuses on **modeling exercises**. While research on activating teaching approaches in software engineering exists [20], other studies discuss best practices for pedagogy and subject-specific teaching of modeling [21]. These practices address issues such as '*tailoring the development process*', '*definition of sequencing UML arte-facts*', '*constant feedback from/to the participants*' and '*con-ducting experiments*'. Authors advocating best practices in software engineering recommend incorporating projects or case studies into courses [22]. The techniques presented in this paper align with these suggested best practices.

In order to focus on exercises, it is essential to consider research on **common mistakes made by students** when working with UML. This paper takes into account up-to-date investigations into frequently occurring errors (refer to [19]). While decision nodes in activity diagrams consistently pose challenges for learners—ranging from '*missing decision node description*' to '*multiple control flows enter one action node*'— loops are not explicitly addressed in the mentioned study. Therefore, alongside the competency-based teaching approach, this paper proposes solutions for modeling and teaching branches and loops in behavioral diagrams.

*2. The Approach of Dealing with Competency-Based Education*

In the context of the German university landscape, there is a prevailing preference for competency-based teaching, as elaborated in the expert report on competency orientation in study and teaching by the German Rectors' Conference (Hochschulrektorenkonferenz - HRK) in 2012 [2]. As a result, numerous degree programs and module handbooks have been designed with a competency-based approach. [3, 4]. Based on Bloom's concepts (1956) [5] a taxonomy matrix has been described at the HRK conference. It was adapted by Anderson and Krathwohl in 2001 [6], and was later expanded once again by HRK in 2013 [3]. This approach follows the "*constructive alignment*" principle (invented by Biggs and Chow [7]).

However, categorizing the approaches presented in this work within the "HRK taxonomy matrix" would depend on the specific objectives of each teaching session. Consequently, only the taxonomy levels' labeling based on Bloom's principles are used, ranging from **remember to create**. This simplifies easy integration into current study programs.

*3. Dealing with Criticism of the Standard Notation*

UML has emerged as the standard for modeling. However, it is essential to recognize that UML does not offer a universal solution to all modeling challenges. Valid criticisms exist regarding this notation. For instance, the modeling of processes in activity diagrams relies on outdated flowcharts (i.e. program flowcharts), often resulting in complex diagrams. This complexity increases when processes contain many nested and repetitive elements. Given UML's status in the modeling domain, I agree that students must be thoroughly prepared to deal effectively with those aspects. Consequently, Chapter III.B discusses some important issues and proposes solutions to enhance proper training for students.

### B. Content of this Work

Chapter II of this paper outlines the essential skills for UML modeling through the six taxonomy levels specified in [3]. The approaches recommended in this paper are shown in Chapter III.

Concrete examples are provided, that are relevant to the practice of teaching. In Chapter VI an evaluation from lectures following the suggested techniques is described. Chapter 0 concludes with a summary and future work.

TABLE I. COMPETENCY CLASSIFICATION IN UML MODELING

| No. | Competencies | Level |
|---|---|---|
| A | Identifying and naming of UML elements | *1-Remember* |
| B | Recognizing diagram types | |
| C | Reading diagrams | *2-Unterstand* |
| D | Inserting, modifying or deleting elements (e.g., multiplicities, stereotypes, visibilities) | *3-Apply* |
| E | Properly applying relationship elements | |
| F | Drawing control structure elements (such as sequence, branching, and loop) | |
| G | Creating comprehensive diagrams | |
| H | Identifying incorrect or allowed elements | *4-Analyze* |
| I | Checking UML syntax for purpose (correct arrow type, correct symbol) | |
| J | Checking order of elements | |
| K | Verifying control structure elements (i.e., sequence, branching, and loop) | |
| L | Analyzing diagrams (e.g. sensibility, meaningfulness, and correctness) | |
| M | Assessing the quality of UML diagrams | |
| N | Evaluating UML diagrams according to problem definition | *5-Evaluate* |
| O | Formulating and achieving modeling goals (To achieve various modeling objectives, one should be familiar with the application and purpose of UML diagram types, trained in their modeling, and capable of utilizing best practices). | *6-Create* |

## II. REQUIRED COMPETENCIES FOR MODELING WITH UML

The proposed approach to teach modeling follow the described six taxonomical levels: **1-remember, 2-under-stand, 3-apply, 4-analyze, 5-evaluate, and 6-create.**

A proper classification of competencies for designing UML diagrams into the suggested taxonomy levels is not a straightforward transaction, so it is advisable for teachers to carefully reconsider such categorization themselves. This paper uses for demonstration the distribution of competencies presented in Table I.

TABLE II. GENERAL AND DIAGRAM-SPECIFIC COMPETENCIES

| General Aspects | |
|---|---|
| (A) Identifying UML-elements | |
| (B) Recognizing diagram types | |
| (C) Reading diagrams | |
| (D) Inserting, modifying, and deleting elements (e.g., multiplicities, stereotypes, visibilities) | |
| (E) Properly applying relationship elements | |
| (F) Drawing control structure elements | |
| (G) Creating a comprehensive diagram | |
| (H) Identifying incorrect/allowed elements | |
| (I) Checking UML syntax for purpose (correct arrow type, correct symbol) | |
| **Specifics of Individual Dynamic Diagram Types:** | |
| *Activity diagram:* | (J) Checking element order |
| | (F) Draw and verify (K) control structure elements (i.e., sequence, branching, and loops) |
| | (L, M) Well-formedness (e.g. merging, or multiple end symbols) |
| | (L, M) Clear labeling |
| *Sequence diagram* | (J) Verify element order |
| | (F, K, L, M) Train consistency between activity and sequence diagrams (advantage: ability of distinguishing between each other) |
| | (F) Practice nesting control structure elements |

In this work, Chapter II demonstrates training aspects, showing how to concretely achieve general (i.e. for all diagram types the same) and diagram-specific competencies listed in Table II. Each list item is aligned with the **uppercase letters** in Table I. Thus, I discuss the competencies outlined in the both tables.

## III. PROPOSED APPROACHES WITH EXAMPLES

### A. General Aspects

The techniques outlined in Table II can be applied to all UML diagram types. A teaching example is provided for each required competence (Table III). Point grading scales (abbr. PGS) can be used, which can be easily applied.

TABLE III. QUESTIONING GENERAL ASPECTS IN UML DIAGRAMS

| Compe-tence | Method Proposal | Level of Taxonomy | Evaluation Option |
|---|---|---|---|
| *Identi-fying UML-elements (A)* | *Knowledge questions* | Remember | PGS |
| | Concrete example: 'A filled diamond is used to represent a composition in a class diagram. (True / False) ', Answer: True. | | |
| *Recogni-zing diagram types(B)* | *Matching questions* | Remember | PGS |
| | To teach learners about the distinctions between diagram types, instructors may use a UML Memory game, either in print, or by utilizing H5P task library [11], or other tools. Concrete example: 'Always look for two matching cards (see Figure 1). ' 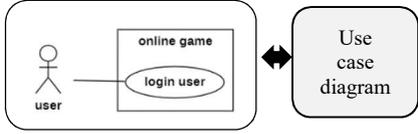 Fig. 1. UML-Memory Game | | |
| *Reading diagrams (C)* | *Questions about image annotations* | Understand | PGS |
| | A useful strategy for enhancing the learners diagram-reading skills is to work with input and output values. This approach offers flexible exercises by using different values and is illustrated through an exercise with a state machine. Concrete example: 'Let us suppose that you enter the number 5 into the 'Number Code Machine' state machine diagram depicted in Figure 2). In this scenario, what letter would the machine produce as its output? ', Answer: A. 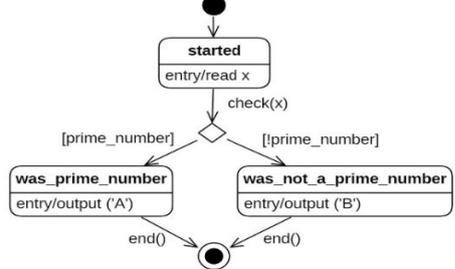 Fig. 2. State machine diagram with input and output | | |
| *Inserting, modifying, and deleting elements (D)* | *Questions about image annotations* | Apply | PGS |
| | Concrete example: 'What should be the multiplicity added to Figure 3 for circular marking 1, and what should be the multiplicity added to marking 2 for an accurate representation? ', Answer: 1- Value '1' and 2- Value 'n'. 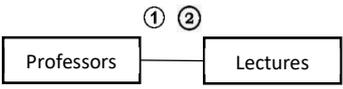 Fig. 3. Question regarding an annotation in an image | | |
| *Properly applying relation-ship elements (E)* | *Questions about image annotations* | Apply | PGS |
| | Using a use case diagram, it is demonstrated how relationship elements and arrow directions can be queried. Concrete example: 'Given is the incomplete UML use case diagram of Figure 4, which is described in the following. Scenario: A teacher administers an exam, which can either be oral or written. For the oral exam, only one student may participate, while any number of students can participate in | | |

| Compe-tence | Method Proposal | Level of Taxonomy | Evaluation Option |
|---|---|---|---|
| | the written exam, but there must be at least one participant. Becoming ill during the exam, students report their illness. To complete the exam, the teacher must record the grades.<br>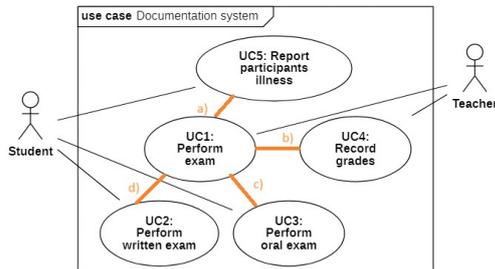<br>Fig. 4. Use case diagram 'documentation system'<br>Question:<br>The type of relationship that the orange line represents at point (a) in the diagram is a/an _______-relationship. The arrowhead of this orange line points to use case number UC-_______.', Answer: 'extend', 'UC-1'.<br>In the digital format, the specific question for the learner can appear as depicted in Figure 5 [10].<br>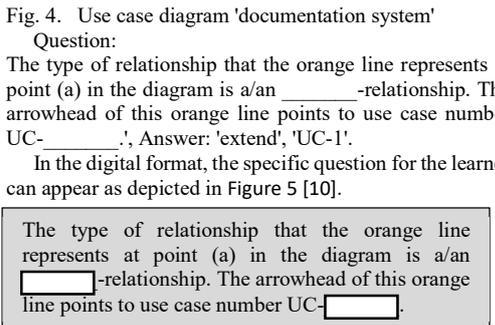<br>Fig. 5. Form of an digital response | | |
| *Drawing control structure elements (F)* | *Creating a comprehensive diagrams* | *Apply* | *PGS* |
| | The practice entails drawing control structure elements, such as sequence, branches, and loops (refer to the discussion in Chapter III.B).<br>Concrete example:<br>'Draw an activity diagram with a while loop',<br>Answer: Please see Figure 6a. Alternatively, a more structured representation with loop nodes can be found in Figure 6b, according to [12].<br>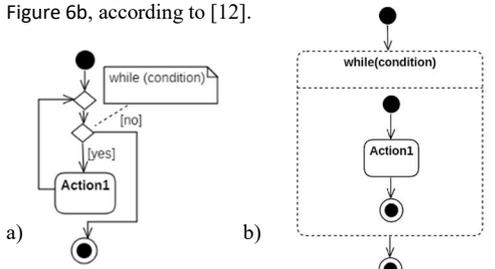<br>Fig. 6. Activity diagram with a while loop | | |
| *Creating a compre-hensive diagram (G)* | *Creating a comprehensive diagram* | *Apply* | *PGS* |
| | For modeling novices, it is advisable to commence with basic, illustrative exercises, whereas professionals may rehearse with intricate, supplied specifications.<br>Concrete example:<br>'Design a use case diagram in which a user has the ability to sign into an online game.', Answer see Figure 7.<br>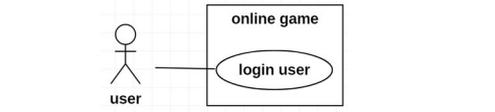<br>Fig. 7. Use case diagram 'online game' | | |
| *Identi-fying incorrect/ allowed elements (H)* | *Finding hotspots in an image* | *Analyze* | *PGS* |
| | With image hotspots, it is possible to inquire about incorrect or allowed elements in diagrams.<br>Concrete example:<br>'In the UML notation, which elements are not permitted in the use case diagram? Please mark/click on the NOT allowed elements in Figure 8',<br>Answer: Both associations between the use cases are not | | |

| Compe-tence | Method Proposal | Level of Taxonomy | Evaluation Option |
|---|---|---|---|
| | permitted. (Associations from the actor to the two lower use cases would need to be added.)<br>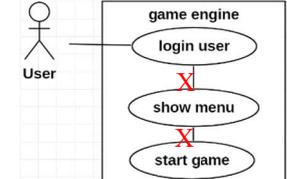<br>Fig. 8. Use case diagram 'game engine' | | |
| *Checking UML syntax for purpose (I)* | *Questions about image annotations* | *Analyze* | *PGS* |
| | Concrete example:<br>'A modeler must depict an asynchronous message in the sequence diagram. Which arrow type should be used (see Figure 9), the type of Message1, or the type of Message2 ?'<br>Answer: Message 2<br>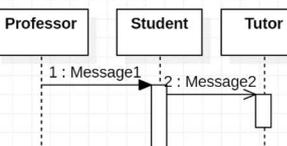<br>Fig. 9. Sequence diagram with an asynchronous message | | |

### B. Specifics of Individual Dynamic Diagram Types

Aspiring modelers must proficiently depict process and program flows, ensuring that the resulting model can be translated into executable software code. Consequently, this work specifically investigates two essential diagram types: activity diagrams and sequence diagrams.

#### 1. Activity Diagrams (F, J, K, L, M)

For a meaningful evaluation of activity diagrams, some guidelines are introduced.

##### a) Guidelines and Reasons for the Guidelines:

As Jacopini showed, in programming, there are three fundamental control structure elements: sequence, branch, and loop [13]. These elements allow modeling of mainly all processes and control flows. As discussed in [15, 23, 14], this work distinguishes between **well-formed** and **abstract** (i.e., non-well-formed) activity diagrams. Well-formed activity diagrams can be easily implemented in programming languages because the guidelines allow for only one end symbol and one token within a single control flow. Exceptions exist in concurrent programs and constructs like 'break' and 'continue.' Loops typically consist of a loop head (diamond symbol), a loop body (activities), and a loop end (merge symbol). An alternative notation using loop nodes and decision nodes is also permissible [12]. Branches begin with a diamond symbol and end with a merge symbol, recognizable in programming languages by the closing parenthesis in if-statements. **Well-formed activity diagrams are closely tied to programming**, while abstract diagrams serve non-programmable processes.

Well-formed activity diagrams are valuable for assessing learners due to **anticipated** solutions and their ability to provide **unambiguous outcomes**. Additionally, they offer a **fair assessment** that can be assigned to all students through a point system. Proficient individuals can **comprehend abstract variations and convert them to well-formed versions**, allowing learners to **actively choose** between the two variants when creating models.

b) *Labeling* and *Commenting*:

Clear and standardized **labels** enhance the readability and learnability of activity diagrams, facilitating **easier corrections**. For branches and loops, a comment field is essential to specify the control structure type (e.g., If, while, do-while, for) and a clear condition (e.g., if (i>1) indicating i is greater than 1). Additionally, numbering branches and loops is mandatory, using the same number in both the opening and closing diamond symbols.

c) *Demonstration*:

For demonstration purposes, this paper provides a straight-forward sample of a properly labeled and well-formed activity diagram, presented in Table IV.

TABLE IV. INQUIRING SPECIFICS OF ACTIVITY DIAGRAMS

| Competence | Method Proposal | Level of Taxonomy | Evaluation Option |
|---|---|---|---|
| *Activity diagram (F, J, K, L, M)* | *Knowledge questions* | *Analyze* | *PGS* |
| | Concrete example: 'Is the activity diagram of Figure 10 well-formed?', | | |

Fig. 10. *Well-formed activity diagram*

Answer: Yes (The well-formed activity diagram presented in Figure 10 features red dashed lines encircling the loop and branching to enhance visibility).

TABLE V. CHECKING WELL-FORMEDNESS IN ACTIVITY DIAGRAMS

| Competence | Method Proposal | Level of Taxonomy | Evaluation Option |
|---|---|---|---|
| *Activity diagram (F, J, K, L, M)* | *Positioning a target on an image* | *Apply / analyze* | *PGS* |

Fig. 11. *NOT-well-formed* activity diagram 'Order' [16]

Concrete example:
'Please arrange the elements provided in Figure 11 correctly! (Note: Elements may be selected zero, one, or multiple times and may also overlap. An incorrect spot must be marked only once with a suitable element. Each element has a target at the top left, which must be precisely aligned with the spot to be marked).',
Answer: see Figure 12.

Fig. 12. Solution for the activity diagram 'Order'.

For practicing well-formed diagrams, abstract variations are useful. The objective is to identify non-well-formed elements or elements missing for well-formedness. (see Table V).

TABLE VI. LEARNING SEQUENCE DIAGRAM - TECHNIQUE 2

| Competence | Method Proposal | Level of Taxonomy | Evaluation Option |
|---|---|---|---|
| *Sequence diagram (F, J, K, L, M)* | *Positioning a target on an image* | *Apply* | *Point grading scale (Evaluation example: One point for each element to be drawn)* |

Fig. 13. Identifying elements in a sequence diagram

Fig. 14. Solution for the sequence diagram in fig. 14

Concrete example - Technique 2:
The use of this question type is appropriate for assessing diagram comprehension and sequencing accuracy.
'Please arrange the elements provided below in Figure 13 correctly!

| Competence | Method Proposal | Level of Taxonomy | Evaluation Option |
|---|---|---|---|
| | | | (Note: Elements may be selected zero, one, or multiple times and may also overlap. An incorrect spot must be marked only once with a suitable element. Each element has a target at the top left, which must be precisely aligned with the spot to be marked). ', Answer: see Figure 14. |

*2. Sequence Diagrams (F, J, K, L, M)*

This work describes two handsome techniques for teaching the characteristics of sequence diagrams.

*a) Technique 1:* The first approach requires that an activity diagram must exactly match a sequence diagram, meaning that the activity diagram must match the sequence diagram one to one. On one hand, this enhances comprehension of the distinctions between activity diagrams and sequence diagrams. On the other hand, nesting of control flow elements such as branches and loops is easier to learn through sequence diagrams since those elements are nested like boxes. Visual detection of incorrect crossovers is facilitated. Exercises can be found in literatur (e.g. [23, 14]).

*b) Technique 2:* The second approach to mastering the fundamentals of sequence diagrams involves studying the precise order of elements and the various possibilities that can occur within these diagrams. (see Table VI).

*C. Advanced Competencies*

Once the basics have been learned to the level of confident use of diagrams (i.e. taxonomy level: apply), advanced studies can begin with training in analysis based on quality criteria (M), evaluation (N), and advanced creation (O). For high taxonomy levels quality assessments of the work are more meaningful, for which grade scales are appropriate.

The suggestion is to switch at these levels to textual questions, such as *'Based on the given requirements and the provided (or self-created) UML diagram, perform an analysis according to the quality criteria of correctness, meaningfulness, and feasibility.'* or to ask for evaluating a model. Regarding advanced diagram creation one could ask *'Create a software architecture for problem X: Formulate modeling goals and demonstrate how you can achieve them with best practices'*. Finally, finding appropriate boundaries between beginner and expert training might be helpful.

## IV. Evaluation

Evaluation method: Observation and time measurement. Before and after introduction to well-formedness in my lectures I asked to write an activity diagram with a branch nested in a loop. Once the approach was adopted, measurements revealed students achieve correct solutions up to three times faster (see table VII), with improved quality (e.g. programmable activity diagrams).

TABLE VII.  APPROACH EVALUATION

| Term | 2022/2023 | 2023 | 2023/2024 |
|---|---|---|---|
| Number of students | 90 | 65 | 100 |
| Time before introduction (min) | 17 | 15 | 19 |
| Time after introduction (min) | 5,5 | 6 | 5 |
| Saved time (%) | 32 | 40 | 26 |
| | | Average (%): | 33 |

## V. Summary and Outlook

This paper outlines a structured approach for students to acquire step-by-step modeling skills. Concrete examples demonstrate how professors can train and assess UML diagrams, whether in digital or written form. The work includes two lists of UML-specific competencies, each aligned with Bloom's taxonomy levels. Concrete teaching exercises illustrate the value of different approaches.

Future work will explore well-formedness in dynamic process models and compare various modeling methods. Additionally, an open-source web-based toolkit is proposed to serve as a repository for UML modeling exercise patterns, fostering collaboration and evidence-based practices.